\documentclass[prl,aps,twocolumn,floatfix]{revtex4}
\usepackage{amsmath,graphics,epsfig,color,verbatim,ulem}

\begin{document}
\title{ Complex Landau Ginzburg Theory of the Hidden Order in URu$_2$Si$_2$}
\author{Kristjan Haule and Gabriel Kotliar}

\affiliation{Center for Materials Theory, Serin Physics
Laboratory, Rutgers University, 136 Frelinghuysen Road,
Piscataway, New Jersey 08854, USA}

\pacs{71.27.+a, 72.15.Rn, 71.30.+h}

\begin{abstract}
We develop a Landau Ginzburg theory of the hidden order phase
and  the local moment antiferromagnetic phase   of URu$_2$Si$_2$.
We  unify  the two broken symmetries in a common complex order
parameter   and derive   many experimentally relevant
consequences such as   the topology of the phase diagram in
magnetic field and pressure.   The theory accounts for  the
appearance of a moment under application of stress and  the
thermal expansion anomaly across the phase transitions. It
identifies  the   low energy mode which is seen in the hidden
order phase near the conmensurate  wavector (0,0, 1) as the
pseudo-Goldstone mode  of the  approximate U(1) symmetry.
\end{abstract}
\maketitle



URu$_2$Si$_2$   is arguably the most intriguing heavy fermion
material and its  electronic structure has continued to be the
focus of intensive investigations.  It displays a  phase
transition at low temperatures material to a phase  for which, in
spite of a large number of experimental efforts,  the order
parameter has not been identified and it is therefore referred to
as hidden order (HO)~\cite{premi_review}.


%
Recent first principles LDA+DMFT calculations showed that in the
paramagnetic phase of URu$_2$Si$_2$, the  local ground state of
the 5f$^2$ configuration, and its first excited state are  two
singlets  separated by a crystal field splitting $\Delta$. In an
atomic picture, 
the corresponding low lying singlets  are:

$|\emptyset\rangle = \frac{i}{\sqrt{2}}(|4\rangle -|-4\rangle)$,
and $|1\rangle = \frac{\cos(\phi)}{\sqrt{2}}(|4\rangle +
|-4\rangle)-\sin(\phi)|0\rangle$,
with $\phi\sim 0.372\pi$ and $\Delta \approx 35 K $.

\begin{figure}[htb]
\centering{
\includegraphics[width=0.7\linewidth]{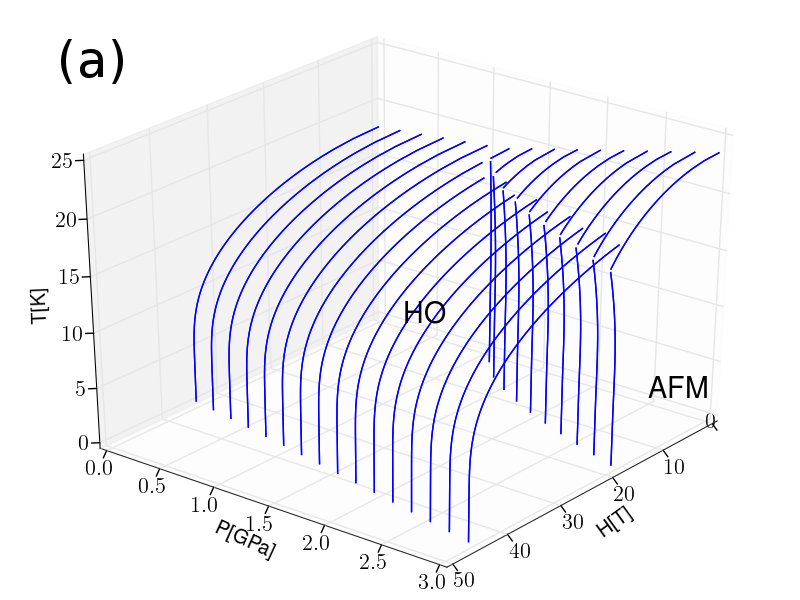}
\includegraphics[width=0.5\linewidth]{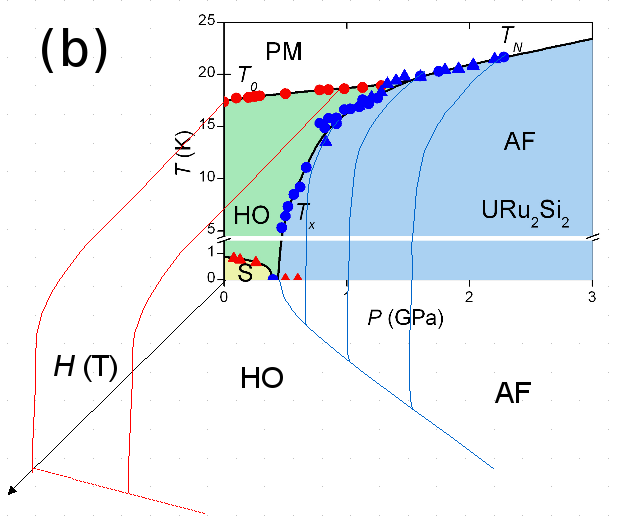}
\includegraphics[width=0.4\linewidth]{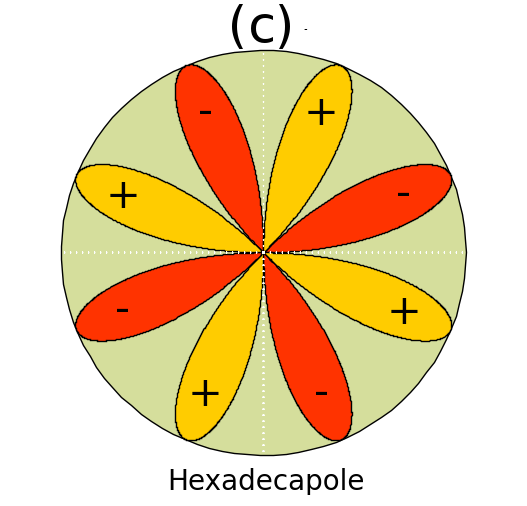}
}
\caption{ (a) The phase diagram of the mean field theory under
  pressure and in magnetic field (b) Experimental phase diagram by
  Aoki~\textit{et al.} \cite{Hassinger}.  (c) The symmetry of the
  hexadecapole order parameter of the HO phase.
}
\label{BP_phased}
\end{figure}

We proposed  that  the order parameter of the hidden order (HO)
and local moment antiferromagnetic (LMA) phase is  the excitonic
mixing between the two lowest lying configurations of the f
electrons ~\cite{previous}. The excitonic mixing of the two
singlets is  described by a complex order parameter $\langle
X_{\emptyset 1}(j)\rangle = \psi_j/2= (\psi_{1,j} + i
\psi_{2,j})/2 $, where $X_{\emptyset 1}(j)$ is the Hubbard
operator $|\emptyset \rangle\langle 1|$ at site $j$. $\psi_2$ is
proportional to the magnetization along the $z$ axis in the
material, while $\psi_1$ is proportional to the hexadecapole
operator $(J_x J_y + J_yJ_x)(J_x^2-J_y^2)$. The characteristic
shape of the hexadecapole is shown in Fig.~\ref{BP_phased}c. The
presence of $\psi_1$ does not break the time reversal symmetry,
nor the tetragonal symmetry, but it does break the reflection
symmetry along the $x$ and $y$ axis. We identify the phase with
nonzero $\psi_1$ as the "hidden order" phase, and the phase with
nonzero $\psi_2$ as the LMA phase. In this picture, the LMA and
the HO order parameters are intimately connected, and they are
related by an internal rotation in parameter space.

In this publication, we build on these insights from LDA+DMFT
microscopic theory to construct a low energy model, and to establish
contact with many of the available experimental results on this
material.
Phenomenological theories of the Landau Ginzburg type for
URu$_2$Si$_2$ have been developed before
\cite{Shah,Mineev,Baek}.
%
However an approximate symmetry between the LMA and and the HO phase
has not been noticed before, and the material specific information
resulting from the microscopic calculations, was not available
before. The new insights from LDA+DMFT calculation restricts the
effective theory enough to result in a large number of consequences
that can be compared with experiment, as for example the response of
the system to pressure, uniaxial stresses and weak external magnetic
field. In addition, the simplifications offered by the low energy
effective Hamiltonian of Landau-Gindzburg type allow us to
obtain analytical expressions which are not possible to obtain in the
full LDA+DMFT solution. 


It is illuminating to interpret the two low lying configurations
of the U atom as
a local two level systems of (pseudo) spins $1/2$ at each lattice
site. The correspondence involves  the three Pauli matrices:
$\sigma_3 =
|\emptyset\rangle\langle\emptyset|-|1\rangle\langle1|$, $\sigma_1
= |\emptyset\rangle\langle1|+|1\rangle\langle\emptyset|$,
$\sigma_2 =
i(|1\rangle\langle\emptyset|-|\emptyset\rangle\langle1|)$.
The crystal field splitting $\Delta$ of the two singlet  states
plays the role of external magnetic field in the third direction,
while the
real and imaginary part of the order parameter appear as the
magnetization along the first and second direction in an internal
order parameter space ($\langle\sigma_1\rangle=\psi_1$,
$\langle\sigma_2\rangle=\psi_2$). The coupling to the external
magnetic field is given by $ B\mu_B(L_z+2 S_z)$ and since only the off-diagonal
terms are nonzero $\langle 1|J_z|\emptyset\rangle=4i\cos(\phi)$ and
they are purely imaginary, the coupling is proportional to $|\langle
1|L_z+2 S_z|\emptyset \rangle|B \sigma_2$.

The coupling between uranium atoms is here modeled by a set of
exchange constants $J^{\alpha}_{ij}$, which we allow to be
different in the two different ordered states $\alpha=1,2$.
Ignoring for the time being the coupling to the fermionic
quasiparticle excitations, the low energy effective  theory can
be described by
\begin{eqnarray}
H &=& \sum_i -\frac{\Delta}{2}\sigma_3^i
-\mu_B B |\langle 1|L_z+2S_z|\emptyset\rangle|\sigma^i_2\nonumber\\
&+&\sum_{i,j} \frac{1}{2}(J^1_{ij}\sigma_1^i\sigma_1^j+J_{ij}^2\sigma_2^i\sigma_2^j).
\end{eqnarray}
%
This Hamiltonian should be regarded as a simplified toy model to
capture symmetry related changes not to far from the phase
transitions.  The LDA+DMFT treatment indicates that a more refined
treatment should include the coupling to dispersing electronic
excitations. This coupling is important for understanding several
properties, including the low tempeature superconductivity and the
fermi surface reconstruction as a result of the coupling of the
fermions to the excitonic mode. The LDA+DMFT Fermi surface has been
shown to give rise to partially nested features with a wavevector
$(0.6,0,0)$\cite{previous}. Consequently, the itinerant electrons
control the mangetic response around this wavevetor as shown in
Ref.~\onlinecite{Janik}. The analysis of the coupling of the mode to
the fermionic quasiparticles is left for future study.

We now treat this Hamiltonian in the mean field approximation and
arrive at an effective free energy of the spin system, written in
terms of the order parameters $\psi_{\alpha,i}$ and the conjugate
Weiss fields $h_{\alpha,i}$ \cite{fukuda}
\begin{eqnarray}
F[h,\psi] = \frac{1}{2} \sum_{ij, \alpha=(1,2)} J_{ij}^\alpha\;  \psi_{\alpha,i}  \psi_{\alpha,j}
 - \sum_{i,\alpha=(1,2)} (h_{\alpha,i} +b_\alpha
 ) \psi_{\alpha,i}\nonumber\\
 -\frac{1}{2}T\sum_i \log\left(\cosh\left(\beta
 \sqrt{\left({\Delta}/{2}\right)^2 + \left({h_{1,i}}\right)^2+
   \left({h_{2,i}}\right)^2}\right)\right).
 \label{landau}
\end{eqnarray}
Here
$b_2 \equiv b= B\mu_B|\langle 1|L_z+2S_z|\emptyset\rangle|$,
with $|\langle 1|L_z+2S_z|\emptyset\rangle|=3.2\cos(\phi)\approx 1.25$, is proportional to the external magnetic
field $B$,  while $b_1$ is a fictitious field which  couples to
the hexadecapole order. The imaginary part of $\psi$ breaks the
time reversal symmetry and therefore couples to the magnetic
field. The mean field equations can be obtained by extremizing
the free energy Eq.~(\ref{landau})
\begin{eqnarray}
h_{\alpha,i}+b_\alpha = \sum_j J_{ij}^\alpha \psi_{\alpha,j}\\
\psi_{\alpha,i} = -\frac{h_{\alpha,i}}{2} \frac{\tanh(\beta\lambda_i)}{\lambda_i}
\end{eqnarray}
with
$\lambda_i = \sqrt{\left({\Delta}/{2}\right)^2 + \left({h_{1,i}}\right)^2+ \left({h_{2,i}}\right)^2}.$

We take the exchange constants between uranium sites in body centered
tetragonal structure to be ferromagnetic in the same plane (RKKY is
ferromagnetic at short distance), and antiferromagnetic in $c$
direction, favoring the staggered order with wave vector $Q=(0,0,1)$,
observed experimentally \cite{Broholm,Broholm2,flauquet0}.
As a result of the close similarity of the exchange constants of the
hexadecapole and the LMA order parameters, both condense at the
ordering vector $Q=(0,0,1)$.

When the restriction to short range exchange is imposed, the only
combination of the exchange constants that enters the mean field
equation is $J_{eff}=4|J_1|+8 J_2$, where $J_1$ is the nearest neighbor
ferromagnetic exchange and $J_2$ is the antiferromagnetic exchange in
$c$ direction. The critical temperature in the absence of magnetic
field is then given by
$T_c=\Delta/(2\;\textrm{atanh}(\Delta/J_{eff}))$.

Near the transition, when the field $\psi$ is small, the free energy
acquires a simple Gindzburg-Landau form
\begin{eqnarray}
F[\psi] \approx \frac{1}{2} \sum_{ij, \alpha=(1,2)} J_{ij}^\alpha\;
\psi_{\alpha,i}  \psi_{\alpha,j}
+\sum_i \widetilde{a} \psi_i^{(2)} + u (\psi_i^{(2)})^2
- b \psi_{2,i}\nonumber
\end{eqnarray}
where
$\psi_i^{(2)} = \sum_\alpha ({\psi_{\alpha,i}})^2$ and
$\widetilde{a}=\frac{\Delta}{2}\coth(\beta\Delta/2)$ while
$u = \frac{\Delta}{8}[\sinh(\beta\Delta)-\beta\Delta]\frac{\cosh^2(\beta\Delta/2)}{\sinh^4(\beta\Delta/2)}$.

We determine the effective exchange constants $J_{eff}$ at zero pressure
in such a way to reproduce the experimentally observed critical
temperatures $J_{eff}^1 = \frac{\Delta}{\tanh(\Delta/(2 T_0))}$ and
$J_{eff}^2 =\frac{\Delta}{\tanh(\Delta/(2 T_N))}$, where $T_0=17.7\,$K
is the hidden order transition and $T_N=15.7\,$K is the Neel
temperature. The exchange constants are strain dependent and for the
small compression we may use the expansion
$$J_{ij}^\alpha \rightarrow
J_{ij}^\alpha(1+g_\alpha(\varepsilon_{xx}+\varepsilon_{yy}))$$ where
$\varepsilon_{xx}$ and $\varepsilon_{yy}$ is the strain (compression)
in $x$ and $y$ direction. The two constants $g_1$ and $g_2$ are
determined in such a way to reproduce the experimental variation of
transition temperature with pressure $T_c(p)$. We take $g_1=20$,
$g_2=49$. 


Having determined all the parameters of the  Landau Ginzburg
functional, we proceed to determine the phase diagram as a
function of pressure and magnetic field, which can be directly
compared to the recent experimental results of
Refs.~\cite{Hassinger,flauquet0}.

There are two solutions of the mean field theory, the local moment
antiferromagnetic solution with nonzero staggered $\psi_{2,i}$ and
vanishing $\psi_{1,i}$. The second solution has nonzero staggered
$\psi_{1,i}$ but vansishing $\psi_{2,i}$, hence it shows no magnetic
moment and corresponds to the hidden order phase.

Figure~\ref{BP_phased}a shows the phase diagram of the Ginzburg-Landau
theory in the magnetic field under applied pressure and
Fig.~\ref{BP_phased}b show the experimentally determined phase diagram
\cite{Hassinger}.
%
At low pressure, with decreasing temperature, there is a second order
phase transition into the hidden order state at temperature
$T_0 = \frac{\Delta \lambda_c}{2\textrm{arctanh}(\Delta \lambda_c/J_{eff}^1)}$
with
$\lambda_c=\sqrt{1+\left(\frac{2 b}{\Delta}\right)^2 \left(\frac{J_{eff}^1}{J_{eff}^1+J_{eff}^2}\right)^2}$.
At a critical field,
$b_c=\frac{\Delta}{2}(J_{eff}^1+J_{eff}^2)\sqrt{1/\Delta^2-1/(J_{eff}^1)^2}$, the hidden
order phase is replaced by the fully polarized paramagnetic phase.

At higher pressures, there is a second order transition into the
antiferromagnetic state at temperature
$T_N = \frac{\Delta\sqrt{1+(2 b)^2}}{2\textrm{arctanh}(\Delta\sqrt{1+(2 b)^2}/J_{eff}^2)}$.
The two phases are separated by a first order boundary shown in
Fig.~\ref{BP_phased}a.
It is also apparent from the figure that magnetic field under high
pressure stabilize the hidden order phase relatively to the
antiferromagnetic phase. The hidden order phase in a magnetic field
develops a uniform magnetization directly proportional to the magnetic
field
$M =B \mu_B^2|\langle
1|L_z+2S_z|\emptyset\rangle|^2|/(J_{eff}^1+J_{eff}^2)\approx
0.01 \mu_B (B/\textrm{T})$, in good quantitative agreement with
experiment of Ref.~\onlinecite{Vojta}.
%
Our model also allows us to extract the jump of the specific heat
across the hidden order to paramagnetic phase boundary.
It is given by
$\Delta c_v =\left(\frac{\Delta}{T_c}\right)^3\frac{1}{4\cosh^2\left(\frac{\Delta}{2T_c}\right)[\sinh(\frac{\Delta}{T_c})-\frac{\Delta}{T_c}]}$,
hence $\Delta c_v/T_c\approx 245 ~\textrm{mJ}/\textrm{mol K}^2$, and
is in good agreement with experiment~\cite{Dijk}.
%
%

\begin{figure}[htb]
\centering{
  \includegraphics[width=0.8\linewidth]{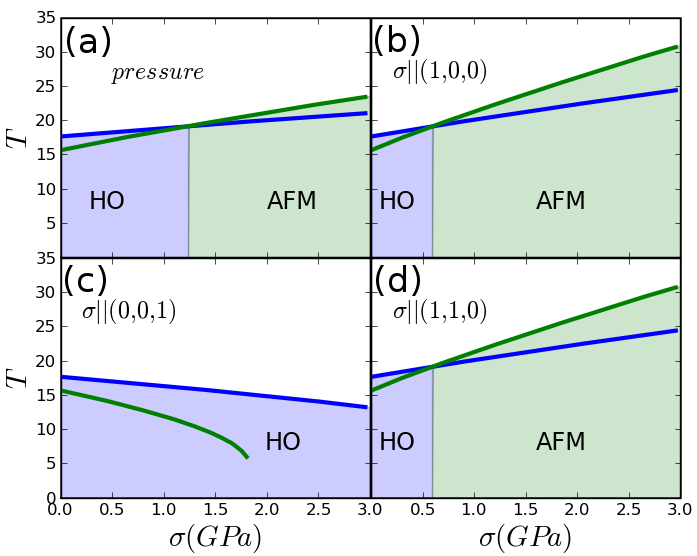}
  }
  \caption{
The phase diagram (a) under pressure and (b),(d) stress in $ab$
plane, and uniaxial strain (c). The critical temperatures below
which the HO and the LMA  solutions are possible are indicated in the
diagram.
The regions where LMA and HO are stable are shaded in blue and
green respectively. The transition between the HO and the LMA
phase is of the  first order. } \label{Fbz}
\end{figure}
The pressure dependence of the critical temperature is shown in
Fig.~\ref{Fbz}a.  The close similarity with experiment of
Hassinger~\textit{et al.}\cite{Hassinger0} is expected, since the
the couplings $g_1$ and $g_2$ were  optimized to reproduce correct
pressure dependence of $T_c$. However, strain in $a$ and $c$
direction acts in quite unexpected way on the stability of the
two phases. The stress is connected to strain by the elastic
constants ($\sigma = C\varepsilon$)(see
Ref.~\onlinecite{Yokoyama} for details). Their value was
determined by the ultrasonic-sound-velocity measurements to be
$(c_{11},c_{33},c_{12},c_{13}) =(255,313,48,86)\times
10^{10}\textrm{erg/cm}^3$ \cite{Wolf}.


Application of strain in $ab$ plane ($\sigma||(1,0,0)$ or
$\sigma||(1,1,0)$) destabilizes the hidden order state, and
stabilizes the antiferromagnetic state at $\sigma$ around
0.6$\,$GPa.
Within our model, the transition temperatures for the two phases are
given by
\begin{eqnarray}
T_{c}^\alpha(\sigma_{xx}) =
\frac{\Delta}{2}\left[\textrm{atanh}\left(\frac{\Delta/2}{J_{eff}^\alpha(1+g_\alpha\frac{c_{33}}{(c_{11}+c_{12})c_{33}-2c_{13}^2}\sigma_{xx})}\right)\right]^{-1}
\nonumber
\end{eqnarray}
Hence LMA is nucleated upon application of very small in-plane
stress. This has been observed in the neutron scattering experiments
of Ref.~\onlinecite{Yokoyama}. In those experiments, the staggered
magnetization was found to grow linearly with the applied stress, an
observation which has been taken as evidence for time reversal
breaking in the hidden order phase~\cite{fazekas-kiss}.  In our
framework, these measurements~\cite{Yokoyama} can be interpreted as
the result of inhomogeneities in the strain field, which given the
very low barrier between HO and the LMA state, easily nucleates LMA
regions. 
The LMA phase persists as a metastable state all the way to zero
strain, indicating that nucleation of LMA regions is actually possible
for infinitesimal stress in the presence of defects. This picture can
be tested by scanning tunnelling microscopy of stressed samples.

So far, we have considered stress breaking the tetragonal
symmetry. Uniaxial strain stabilizes hidden order (the T$_{Neel}$
smaller than $T_0$) and decreases the hidden order transition
temperature, as shown in Fig.~\ref{Fbz}c.
The formula for the two transition temperatures is given by
\begin{eqnarray}
T_{c}^\alpha(\sigma_{zz}) =
\frac{\Delta}{2}\left[\textrm{atanh}\left(\frac{\Delta/2}{J_{eff}^\alpha(1-g_\alpha\frac{2c_{13}}{(c_{11}+c_{12})c_{33}-2c_{13}^2} \sigma_{zz})}\right)\right]^{-1}.
\nonumber
\end{eqnarray}
The $5.6\,$GPa strain leads to complete elimination of the hidden
order. However, the uniaxial strain   also affects the crystal
field splitting $\Delta$.    We have not attempted to model this
dependence quantitatively in our theory, but we notice that an
increase in $\Delta$  will result in a rapid decrease of  the
hidden order  temperature (and the LMA critical temperature)
resulting in a quantum critical point at  critical value of the
crystal field splitting  $\Delta_c = {J^1}_{eff}$ separating the
HO phase from the paramagnetic phase at zero temperature. This
critical point may already have been accessed via Rh doping in
Ref.~\onlinecite{RhDop1}. This quantum critical point is expected
to be in the same universality class as that of Ising
itinerant antiferromagnets \cite{millis}. 

\begin{figure}[htb]
\centering{
  \includegraphics[width=0.8\linewidth]{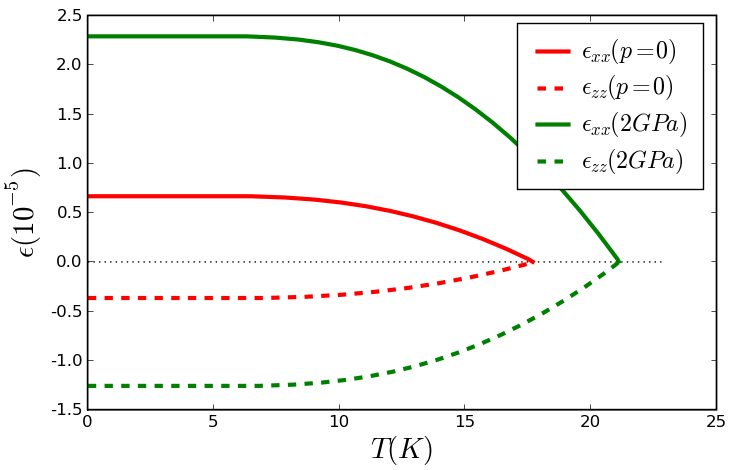}
  }
  \caption{
The strain resulting from  the  phase transition into HO and LAM
state. The sample contracts in $ab$ plane (positive strain), and
expands in $c$ direction (negative strain). } \label{Lbz}
\end{figure}

To study the dilatation of the sample in $a$ and $c$ direction
due to the emergence  of the HO and LMA states we  add to the
Landau free  energy the standard elasticity terms  and their
coupling to the order parameter

\begin{eqnarray}
\frac{1}{2}\varepsilon C \varepsilon - \varepsilon \sigma +
\frac{1}{2} \sum_{ij, \alpha=(1,2)}
J_{ij}^\alpha(1+g_\alpha(\varepsilon_{xx}+\varepsilon_{yy}))\;
\psi_{\alpha,i}  \psi_{\alpha,j}. \nonumber
\end{eqnarray}

Here $\sigma$ and $\epsilon$ are the stress and strain tensor and
the tensor C describe the elastic moduli. Differentiating this
free energy we obtain  the dilatation in $a$ and $c$ direction,
$\varepsilon_{xx} =c_{33}L_\alpha/2$ and $\varepsilon_{zz}
=-c_{13} L_\alpha$, where $L_\alpha = \frac{g_\alpha
J_{eff}^\alpha
(\psi_{\alpha,i})^2}{((c_{11}+c_{12})c_{33}-2c_{13}^2)}$.

%

Fig.~\ref{Lbz} shows the temperature dependence of the strain in
the $a$ and $c$ direction at zero pressure, where the transition
into the hidden order state occurs. The sample contracts in the
$ab$-plane (positive strain) and expands in $c$ direction
(negative strain). The expansion in $c$ direction is smaller than
expansion in $ab$ direction. At  a pressure of $2\,$GPa, where the
LMA state is stable, the dilatation has the same trend but the
magnitudes are  considerably larger. 
Similar temperature variation of the dilatation was recently
measured in Ref.~\onlinecite{dilatation}, where considerably
larger dilatation was found in transition to LMA phase than
to the hidden order phase. This difference is connected to the larger
slope of the LMA  transition temperature compared to hidden order
transition temperature.





The Landau Ginzburg free energy explicitly exhibits the remarkable
similarity between the hidden order phase and LMA phase. There is an
approximate U(1) rotational symmetry in an internal parameter space.
The main difference between the two phases lies
in the different values of the exchange constants, which is of the
order of 6$\,\%$. Our theory therefore provides a microscopic basis
for the remarkable similarity between the LMA and HO phase, dubbed
\textit{adiabatic continuity} \cite{adiabaticc}.


If the U(1) symmetry was exact, the spontaneous breaking of this
symmetry would result in a Goldstone mode describing the transverse
fluctuations of the order parameter, i.e., the potential would have
the form of a Mexican hat with a flat bottom. Due to the small
explicit breaking of the U(1) symmetry, since the exchange constants
$J_{eff}^1$ and $J_{eff}^2$ are slightly different, this Goldstone
mode acquires a small mass, and we refer to it as a pseudo-Goldstone
mode.  In the hidden order phase $\langle\psi_{1,i}\rangle \neq 0$ the
pseudo-Goldsone mode can be identified with $\psi_{2,i}$ (the
transverse fluctuation in the Mexican hat).  Hence the
pseudo-Goldstone mode of the hidden order phase carries a magnetic
moment and can be observed by neutron scattering. In the
antiferromagnetic phase the pseudo-Goldstone mode can be identify with
$\psi_{1,i}$, which carries hexadecapolar moment but no mangetic
moment and is invisible to neutrons. As a result a low lying mode at
$(0,0,1)$ is only visible to neutrons in the hidden order phase.  This
provides a natural explanation of the mode observed in neutron
scattering experiments, as measured by Broholm \textit{et
  al.}~\cite{Broholm,Broholm2} and Villaume~\textit{et
  all.}~\cite{flauquet0}. The mode was observed only in the hidden
order phase, but not in the antiferromagnetic phase~\cite{flauquet0}.
The energy scale of this mode is a measure of how different the hidden
order phase is from the antiferromagnetic phase, and therefore should
decrease with pressure (since the difference in the exchange constants
decreases with increasing pressure) and should increase with magnetic
field (since the magnetic field destabilizes the antiferromagnetic
phase further relative to the hidden order phase).

In conclusion, we developed an effective theory of the
paramagnetic to hidden order and local antiferromagnetic
transition. The theory is consistent with a large body of
experimental data, is inspired by microscopic LDA+DMFT
calculations, and puts URu$_2$Si$_2$ in a broader context of
other $f^2 $ systems \cite{flouquet,takimoto}.

Acknowledgement: We are grateful to P. Chandra and P. Coleman for
a useful discussion. We are particularly grateful to D. Aoki, E.
Hassinger, and J. Flouquet for permission to reproduce their
experimental phase diagram in our figure.
 K.H was supported by Grant
NSF NFS DMR-0746395 and Alfred P. Sloan fellowship. G.K. was
supported by NSF DMR-0906943.

%
%
%
%
%
%
%
%
%
%
%
%
%
%
%
%
%
%
%
%
%
%
%
%
%
%
%
%
%
%
%
%
%
%
%
%
%
%
%
%
%
%
%
%

\end{document}